\documentclass[11pt]{article}
\pdfoutput=1
\usepackage{jcappub,natbib,float,comment,bm,placeins} 
\usepackage{soul}

\def\be{\begin{equation}}
\def\ee{\end{equation}}
\def\ba#1\ea{\begin{align}#1\end{align}}

\newcommand{\bq}{\begin{eqnarray}}
\newcommand{\eq}{\end{eqnarray}}

\renewcommand{\v}[1]{\mathbf{#1}}
\newcommand{\vx}{\v{x}}

\newcommand{\vk}{\v{k}}

\newcommand{\refeq}[1]{Eq.~(\ref{eq:#1})}          
          
\newcommand{\reffig}[1]{figure~\ref{fig:#1}} 
\newcommand{\reffigs}[2]{figures~\ref{fig:#1}--\ref{fig:#2}}
\newcommand{\refFig}[1]{Figure~\ref{fig:#1}}
          
\newcommand{\refsec}[1]{section~\ref{sec:#1}}          
\newcommand{\refSec}[1]{Section~\ref{sec:#1}}

\renewcommand{\d}{\delta}

\def\iMpch{\,h\,{\rm Mpc}^{-1}}

\newcommand{\eps}{\epsilon}

\def\O{\mathcal{O}}
\def\P{\mathcal{P}}
\def\Mpch{\,h^{-1}{\rm Mpc}}

\newcommand{\bkk}{b_{K^2}}

\title{Assembly bias in quadratic bias parameters of dark matter halos from forward modeling}

\author[a,b,c]{Titouan Lazeyras,}
\author[d,e]{Alexandre Barreira,}
\author[f]{Fabian Schmidt}
\affiliation[a]{SISSA, Via Bonomea 265, 34136 Trieste, Italy}
\affiliation[b]{INFN, Sezione di Trieste, Via Bonomea 265, 34136 Trieste, Italy}
\affiliation[c]{IFPU, Institute for Fundamental Physics of the Universe, via Beirut 2, 34151, Trieste, Italy}
\affiliation[d]{Excellence Cluster ORIGINS, Boltzmannstraße 2, 85748 Garching, Germany}
\affiliation[e]{Ludwig-Maximilians-Universit\"at, Schellingstraße 4, 80799 M\"unchen, Germany}
\affiliation[f]{Max–Planck–Institut f\"ur Astrophysik, Karl–Schwarzschild–Strasse 1, 85748 Garching, Germany}
\emailAdd{tlazeyra@sissa.it, alex.barreira@origins-cluster.de, fabians@mpa-garching.mpg.de}

\abstract{We use the forward modeling approach to galaxy clustering combined with the likelihood from the effective-field theory of large-scale structure to measure assembly bias, i.e.~the dependence of halo bias on properties beyond the total mass, in the linear ($b_1$) and second order bias parameters ($b_2$ and $\bkk$) of dark matter halos in $N$-body simulations. This is the first time that assembly bias in the tidal bias parameter $\bkk$ is measured. We focus on three standard halo properties: the concentration $c$, spin $\lambda$, and sphericity $s$, for which we find an assembly bias signal in $\bkk$ that is opposite to that in $b_1$. Specifically, at fixed mass, halos that get more (less) positively biased in $b_1$, get less (more) negatively biased in $\bkk$. We also investigate the impact of assembly bias on the $b_2(b_1)$ and $\bkk(b_1)$ relations, and find that while the $b_2(b_1)$ relation stays roughly unchanged, assembly bias strongly impacts the $\bkk(b_1)$ relation. This impact likely extends also to the corresponding relation for galaxies, which motivates future studies to design better priors on  $\bkk(b_1)$ for use in cosmological constraints from galaxy clustering data.}

\begin{document}
\maketitle
\flushbottom

%%%%%%%%%%%%%%%%%%%%%%%%%%%%%%%%%%%%%%%%%%%%%%%%%%%%%%%%%%%%%%%%%%%%%%%%%%%%%
%%%%%%%%%%%%%%%%%%%%%%%%%%%%%%%%%%%%%%%%%%%%%%%%%%%%%%%%%%%%%%%%%%%%%%%%%%%%%
\section{Introduction}
\label{sec:intro}
 
It is now well established that most of the observed tracers of the large-scale structure (LSS) of the Universe, such as galaxies, reside in dark matter halos. Hence the statistics of halos help determine those of galaxies on large scales, making their distribution one of the key ingredients of the theoretical description of LSS. In the context of perturbation theory, the statistics of halos can be written in terms of bias parameters $b_\O$ multiplying operators $\O$ constructed out of the large-scale matter density contrast field $\d_m$ and tidal field $K_{ij}$ (see \cite{Desjacques:2016} for a recent review)
\be
\d_h(\vx,z) = \sum_\O b_\O(z) \O(\vx,z) \,,
\label{eq:biasexp}
\ee
where $\delta_h$ is the density contrast of the halo distribution, defined as $\delta_h=n_h/\bar{n}_h-1$ with $\bar{n}_h$ the cosmic halo number density. This bias expansion formalism can be applied to any LSS tracer, with only the numerical value of the bias parameters changing from tracer to tracer. In this work we focus on dark matter halos identified in numerical $N$-body simulations. Physically, the halo bias parameters describe the response of the halo number density field to the long-wavelength perturbations $\O$, and have been measured in multiple works at various orders in perturbation theory (e.g.~\cite{Tinker:2010my,2012PhRvD..85h3509C,Sheth:2012fc,Saito:2014qha,Lazeyras:2015,Hoffmann:2016omy,Desjacques:2016,Lazeyras:2017,Abidi:2018,Oddo:2019run,Eggemeier:2021cam,Barreira:2021ukk} and references therein). Studying them can thus give us important insights on halo, galaxy, and in general structure formation. A robust knowledge of halo bias also helps guide the design of priors on bias parameters that enter theoretical models in LSS analyses, or even reduce the number of free parameters if useful relations between them can be found (e.g.~\cite{Lazeyras:2015,2016arXiv160200674B,Hoffmann:2016omy, Oddo:2019run,2020PhRvD.102l3521W,Eggemeier:2021cam}), thereby leading to improved constraints on cosmological parameters.

The bias parameters were commonly thought to depend only on the redshift and mass of the considered halo population. The authors of \cite{Sheth:2004} were the first to show that the formation of halos of the same mass depends also on their environment. Shortly after, \cite{gao:2005} presented the first direct measurements of the dependence of halo bias on properties beyond their mass, showing that halos with a lower half-mass assembly redshift are less biased than average, and vice-versa. This phenomenon was naturally named halo \textit{assembly bias}. Since then, numerous other works, using numerical simulations, have studied and found halo assembly bias as a function of various halo properties, including formation time, concentration, spin, shape, substructure content and mass accretion rate (see for instance \cite{Sheth:2004, gao:2005, Gao:2006, Wechsler:2005, Jing:2006, Croton:2006, Angulo:2007, Dalal:2008, Faltenbacher:2009, Reid:2010,Lacerna:2012, Sunayama:2015, Paranjape:2016, Lazeyras:2016, Salcedo:2017, Mao:2017, Chue:2018, Sato-Polito:2018, Paco_18,Lazeyras:2020suj,Contreras:2021oxg} and references therein). Note that in the literature some works refer to the dependence of halo bias on these secondary halo properties as {\it secondary bias}, reserving the term assembly bias for the dependence of halo bias specifically on the assembly history (i.e.~mass accretion rate or formation redshift). We do not make this distinction here, and from hereon, when we refer to assembly bias, we mean the dependence of the halo bias parameters on properties beyond halo mass and redshift.

Despite simplified analytical approaches based on \textit{excursion set} formalisms (e.g.~\cite{Lazeyras:2016,Borzyszkowski:2016,Musso:2017}), a complete, first-principled theoretical description of assembly bias with respect to any halo property over a wide mass range is still lacking to this day, and most of the recent progress has come from analyses of numerical $N$-body simulations. It is now clear that the halo environment, characterized by its position in the cosmic web, and surrounding density and tidal fields, plays a crucial role in assembly bias \cite{Dalal:2008,Zhang:2013lda,Borzyszkowski:2016, Lee:2016, Lazeyras:2016,Tojeiro:2016nmy,Castorina:2016, Musso:2017, Han:2018, Mansfield:2019, Ramakrishnan:2019}, which motivates the study of assembly bias in tidal bias parameters such as $\bkk$. It is also established that correlations between any two halo properties are not sufficient to predict the dependence of halo bias on one variable given its dependence on the other \cite{Croton:2006,Lazeyras:2016, Mao:2017}.

At fixed halo mass, it is plausible that certain galaxy populations may preferentially inhabit halo populations with specific values for properties like concentration, formation time or environment. The exact degree to which the halo assembly bias signal propagates to galaxy assembly bias remains however an open question, and it likely depends on the details of galaxy formation. Some works have studied this using hydrodynamical simulations, and have reported a detection of assembly bias in the simulated galaxy density field as well \cite{Chaves-Montero:2015,Montero-Dorta:2020tpt,Montero-Dorta:2021uid}. These results have led to several recent efforts to incorporate it in semi-analytical models such as (decorated) HOD  and abundance matching techniques\footnote{Notice that in theory the complete perturbative bias expansion automatically takes into account assembly bias if all necessary terms are included (see section~9.2 of \cite{Desjacques:2016}), but these methods have the advantage to allow to extract information from the non-perturbative regime, where perturbation theory fails.} \cite{Wang:2013,Zentner:2013pha,Pujol:2014avn,Hearin:2015, Chaves-Montero:2015, Lehmann:2015,McEwen:2016obj,Zehavi:2017,Jimenez:2020kra, Zehavi:2019, Contreras:2020, Xu:2020,Contreras:2021oxg}. On the observational front, however, the detection of assembly bias in observational data remains more debated: despite a number of studies claiming a detection (e.g.~\cite{Yang:2005, Lacerna:2011, Tinker:2012, Wang:2013,Hearin:2014uia,Miyatake:2015, More:2016,Obuljen:2020,Yuan:2020xlk,Jimenez:2020kra}), re-analyses and other independent investigations using similar galaxy samples have returned no clear evidence of assembly bias in observations (e.g.~\cite{Lin:2015,Zu:2016,Vakili:2016kkk,Busch:2017,Dvornik:2017vfq,Niemiec:2018lpn,Sunayama:2019,Yuan:2019fbi,Salcedo:2020mar}).

In this paper we take further steps in the study of assembly bias by measuring its impact on the two quadratic bias parameters $b_2$ and $\bkk$, which correspond to the second-order response of halo abundance to large-scale matter overdensities, and to the leading-order response to large-scale tidal fields, respectively. Although the assembly bias signal in $b_2$ has been analyzed in past works  \cite{Angulo:2007,Lazeyras:2016}, this is the first time, to the best of our knowledge, that assembly bias in the halo tidal bias parameter $\bkk$ is estimated. We carry out our analysis using the \textit{forward modeling} approach to halo clustering combined with a likelihood derived from the \textit{effective-field theory} (EFT) of LSS \cite{2019JCAP...01..042S,2020JCAP...04..042C,2020JCAP...07..051C}. This approach has been developed recently, but is related to the analyses presented in \cite{Lazeyras:2017,Abidi:2018,Schmittfull:2019,Schmittfull:2020}. Previous studies showed that the EFT likelihood is able to recover unbiased cosmological constraints using simulated data, even after marginalizing over a large set of bias parameters \cite{2019JCAP...01..042S, 2020JCAP...01..029E, 2020JCAP...11..008S, 2020arXiv200914176S, 2020arXiv201209837S, 2021JCAP...03..058N, Barreira:2021ukk}. In this work, we apply it instead to the study of the bias parameters themselves. We refer the interested reader also to our companion paper \cite{Barreira:2021ukk}, where we applied the same forward model methodology to study the quadratic bias parameters of galaxies in hydrodynamical simulations.  

This paper is organized as follows: in \refsec{biasfwd}, we review the main ingredients of the forward modeling methodology that we use in this paper. We present our simulation set as well as the construction of the halo catalogs in \refsec{sims}. In \refSec{results}, we present and discuss our main results, before we conclude in \refsec{concl}. 

%%%%%%%%%%%%%%%%%%%%%%%%%%%%%%%%%%%%%%%%%%%%%%%%%%%%%%%%%%%%%%%%%%%%%%%%%%%%%
%%%%%%%%%%%%%%%%%%%%%%%%%%%%%%%%%%%%%%%%%%%%%%%%%%%%%%%%%%%%%%%%%%%%%%%%%%%%%

\section{Bias parameters from forward modeling}
\label{sec:biasfwd}

In this section, we summarize the main aspects of our approach to study halo bias using forward models and the EFT of LSS. This methodology has been presented and tested in detail in a number of recent papers \cite{2019JCAP...01..042S, 2020JCAP...01..029E, 2020JCAP...04..042C, 2020JCAP...07..051C, 2020JCAP...11..008S, 2020arXiv200914176S, 2020arXiv201209837S}, to which we refer the reader for more details. The methodology we follow here is largely the same as in our companion paper \cite{Barreira:2021ukk} focused on the bias of galaxies in hydrodynamical simulations, except that the larger volume of the simulations we employ in this paper allows us to carry out a finer investigation of the impact of the range of scales used in the analysis (see \refsec{cutoff} below).

\subsection{Forward modeling and the EFT likelihood}
\label{sec:fwdmodel}

In the forward modeling approach the goal is to derive the \emph{likelihood} function $\P\big(\delta_h | \{\theta\}, \{b_\O\}, \delta_{m,\rm in}\big)$ that describes the probability to observe a halo density contrast field $\delta_h$ (or any other tracer field), given a realization $\delta_{m,\rm in}$ of the initial matter density contrast, a set of cosmological parameters $\{\theta\}$ and a set of bias parameters $\{b_\O\}$. This can be done as follows:
\begin{enumerate}

\item In a given cosmology, evolve $\delta_{m,\rm in}$ under the action of gravity to generate the forward-evolved final matter distribution, $\delta_{m,\rm fwd}[\{\theta\}, \delta_{m,\rm in}]$. In this paper, we do so using third-order Lagrangian perturbation theory (3LPT) \cite{2020arXiv201209837S}.

\item Generate the forward-evolved halo distribution using the final matter distribution and a deterministic bias expansion, $\delta_{h, \rm det}[\delta_{m, \rm fwd}, \{b_\O\}]$.

\item Sample the likelihood $\P\big(\delta_h | \delta_{h, {\rm det}}\big) \equiv \P\big(\delta_h | \{\theta\}, \{b_\O\}, \delta_{m,\rm in}\big)$ in the combined parameter space consisting of the initial conditions $\delta_{m, \rm in}$, the cosmological parameters $\{\theta\}$ and the bias parameters $\{b_\O\}$.

\end{enumerate}
In this paper we consider dark matter halos from numerical gravity-only $N$-body simulations, which allows us to keep the cosmological parameters $\{\theta\}$ and initial conditions field $\delta_{m, \rm in}$ fixed to those used to run the simulations. This drastically reduces the parameter space that would otherwise have to be explored, as well as the impact of sample variance, and thus, we can fit more efficiently for the bias parameters $b_\O$. That is, instead of comparing the measured summary statistics of halos, such as power spectrum and bispectrum, with the theoretically predicted ensemble mean of these statistics, we compare directly at the field level for the given specific realization of the large-scale perturbations, reducing cosmic variance to the highest degree possible.

We work with the following likelihood function in Fourier space
\bq\label{eq:eftlike}
-2{\rm ln}\P\big(\d_h | \d_{h, {\rm det}}\big) = \int_{|\vk| < \Lambda} \frac{{\rm d}^3\vk}{(2\pi)^3} \Bigg[\frac{\big|\d_{h}(\vk) - \d_{h, {\rm det}}(\vk)\big|^2}{P_\eps(k)} + {\rm ln}\big(2\pi P_\eps(k)\big)\Bigg],
\eq
which has been derived using EFT in \cite{2019JCAP...01..042S, 2020JCAP...01..029E, 2020JCAP...04..042C, 2020JCAP...11..008S, 2020JCAP...07..051C, 2020arXiv200914176S} (we distinguish Fourier- from configuration-space quantities by their arguments). The variance $P_\eps(k)$ in this equation takes into account the stochasticity of halo formation (i.e.~its dependence on the small-scale perturbations), and it is in general a function of wavenumber \cite{2020JCAP...04..042C, 2020JCAP...07..051C} (note that \cite{hamaus/etal:2010} has used a similar likelihood function to study halo stochasticity). For our purpose in this paper, it is sufficient to consider only the leading order, constant contribution $P_\eps(k) \approx P_\eps^{\{0\}}$, where $P_\eps^{\{0\}}$ is a parameter that is also sampled (see \cite{2020JCAP...11..008S, 2020arXiv200914176S} for a justification of the small impact of this approximation). 

It is important to note that the integral in Eq.~(\ref{eq:eftlike}) is performed only up to a maximum \textit{cutoff} wavenumber $\Lambda$. This cutoff is needed to ensure that only perturbative modes enter in the inference analyses (in accordance with the EFT of LSS formalism), and as discussed in \cite{2020JCAP...11..008S, 2020arXiv200914176S}, it must be applied also to the initial conditions field $\delta_{m, \rm in}$ before evolving it to the final time in order to regularize loop integrals involving non-perturbative modes. Further, the integral in Eq.~(\ref{eq:eftlike}) is in practice replaced by a sum over the nodes of a regular, cubic grid covering the simulation volume onto which all fields are discretized. Both matter and halos were deposited on grids using the cloud-in-cell (CIC) assignment scheme, and all grids we use here have $N_{\rm grid} = 768$ nodes on a side. This corresponds to a resolution of $L_{\rm cell}=3.125 \Mpch$, and associated Nyquist wavenumber $k \approx 1 \iMpch$, safely above all the wavenumbers considered in the analysis below. We further tested the robustness of our results under a change in $N_{\rm grid}$.

Effectively, the estimation of bias parameters in this approach is closely related to that based on the cross-correlation of the halo field with quadratic operators constructed from the linear density field of the same realization presented in \cite{Lazeyras:2017,Abidi:2018}. This correspondence can be seen in particular by inspecting the maximum-a-posteriori point derived in \cite{2019JCAP...01..042S,2020JCAP...11..008S}.

%%%%%%%%%%%%%%%%%%%%%%%%%%%%%%%%%%%%%%%%%%%%%%%%%%%%%%%%%%%%%%%%%%%%%%%%%%%%%

\subsection{The bias expansion}
\label{sec:biasexp}

The forward-evolved halo density field is constructed out of the forward-evolved matter density contrast $\delta_m \equiv \delta_{m,\rm fwd}(\vx)$ using the deterministic bias expansion
\bq\label{eq:biasexp_E}
\d_{h, {\rm det}}(\vx) = \sum_\O b_\O \O[\delta_{m,\Lambda}](\vx),
\eq
where the notation highlights that the operators are constructed out of the evolved density field filtered with a sharp-$k$, low-pass filter. Notice that the shape of the filter is not arbitrary and that a sharp-$k$ filter is necessary to filter out all modes $k > \Lambda$ and obtain unbiased results from forward models. This is discussed in detail in section 4.3 of \cite{2019JCAP...01..042S}.

We consider the following set of 8 operators
\bq\label{eq:operators}
\O \in \big\{\delta_m, \delta_m^2, K^2, \delta_m^3, \delta_m K^2, K^3, O_{\rm td}, \nabla^2\delta_m\big\},
\eq
where $K_{ij} = \big(\partial_i\partial_j/\nabla^2 - \delta_{ij}/3\big)\delta_m$, $K^2 = K_{ij}K^{ij}$, $K^3 = K_{ij}K^{jk}K_k^i$ and $O_{\rm td} = (8/21) K_{ij} (\partial_i\partial_j/\nabla^2)$ $\big( \delta_m^2 - (3/2)K^2 \big)$. The first 7 represent all terms that exist up to third order in $\delta_m$ and $K_{ij}$. Additionally, there are also higher-derivative operators that should be taken into account. According to the strategy presented in \cite{2020arXiv200914176S}, the relevance of the higher-derivative terms can be estimated using the nonlinear scale $k_{\rm NL}$, the cutoff $\Lambda$, and the spatial nonlocality scale of the tracers $R_*$. In this estimate we adopt $R_* = 5{\rm Mpc}/h$ (which is approximately the Lagrangian radius of the halos we consider), $\Lambda = 0.14 h/{\rm Mpc}$ and $k_{\rm NL} = 0.25 h/{\rm Mpc}$. For these scaling parameters, $\nabla^2\delta_m$ is the only relevant higher-derivative operator (eighth term in \refeq{operators}). Note that we keep these three scaling parameters fixed for the purpose of determining how many higher-derivative operators we need to include in the bias expansion, so that the set of operators remains the same when we vary the redshift and the exact values of $\Lambda$ in our analysis below. We refer the reader to section 3.2 of \cite{2020arXiv200914176S} for a detailed description of how to determine which higher-derivative parameters to include, and the exact combination of  $k_{\rm NL}$, $\Lambda$, and $R_*$ that needs to be computed. Finally, notice that we will use the notation $b_n \equiv n! \, b_{\d^n}$ ($n \geq 1$) for the bias parameters multiplying powers of $\delta_m$ in order to conform to the most common notation in the literature for these parameters.

The operators $\delta_m^2, K^2$ are further renormalized with respect to $\delta_m$ by subtracting their overlap with $\delta_m$ as described in \cite{2019JCAP...01..042S}. This is important to be able to interpret the corresponding bias parameters as those that appear in the large-scale limit of $N$-point functions \cite{2019JCAP...01..042S}. We do not perform renormalization involving the quadratic operators themselves, which most likely explains part of the dependence of our results on the cutoff scale $\Lambda$; we return to this issue in \refsec{cutoff}.

In summary, the construction of $\d_{h, {\rm det}}(\vk)$ is done as follows. The initial field $\delta_{m, \rm in}(\vx)$, which in our case is the Zel'dovich field at $z=99$, is discretized on a grid and filtered with a sharp-$k$, low-pass window function on the scale $\Lambda$ (this is done by going back-and-forth between Fourier and configuration space). This field is evolved to the final redshift with 3LPT (which \cite{2020arXiv201209837S} showed is accurate enough on the scales we consider) to obtain the final mass distribution $\delta_{m}(\vx)$, which is itself subsequently sharp-$k$ filtered on the scale $\Lambda$. The final filtered field is used to construct the operators $\O[\delta_{m,\Lambda}]$ in Eq.~(\ref{eq:operators}) (again by by going back-and-forth between Fourier and configuration space), which are added up as in Eq.~(\ref{eq:biasexp_E}) to generate the forward-evolved halo density field $\d_{h, {\rm det}}(\vx)$. Finally, this field and the {\it observed} halo sample $\d_h(\vx)$ (i.e., the halos in our simulations) are transformed to Fourier space and used in the EFT likelihood of Eq.~(\ref{eq:eftlike}). Notice that the observed halo density field is not filtered, but that this is not needed since the sum in \refeq{eftlike} runs only over modes $k$ smaller than the cutoff $\Lambda$ which is the same used to filter the matter fields from which we construct the bias expansion.

%%%%%%%%%%%%%%%%%%%%%%%%%%%%%%%%%%%%%%%%%%%%%%%%%%%%%%%%%%%%%%%%%%%%%%%%%%%%%

\subsection{The fitting procedure}
\label{sec:fitting}

To fit for a bias parameter $b_\O$, we first marginalize over all others in Eq.~(\ref{eq:eftlike}) (which can be done analytically \cite{2020JCAP...01..029E,2020JCAP...11..008S}), and then find the maximum of the likelihood in the $\{b_\O, P_\eps^{\{0\}}\}$-space (using the {\sc minuit} routine from the {\sc root}\footnote{https://root.cern.ch/} package). Our inferred bias parameter values correspond to the maximum-likelihood values of $b_\O$; the corresponding errorbars are the inverse square root of the curvature of $-2{\rm ln}\P$ in the $b_\O$ direction at the maximum. Formally, in this procedure one should marginalize also over $P_\eps^{\{0\}}$, but since $P_\eps^{\{0\}}$ and the $b_\O$ are only very weakly correlated, this does not have a strong impact on our quoted errorbars. Recall also that in our fitting procedure we always keep the initial conditions, as well as all cosmological parameters fixed to the values of our simulations.

Although the minimization procedure is formally less stable than a proper sampling of the full shape of the likelihood and may sometimes converge to a local (and not the global) minimum, we will see in our results below that this does not have a strong impact on our results as the forward models correctly recover the bias parameters measured in past works. Finally, even though the code also returns estimates of the remaining, higher-order bias parameters, we will show results only for the bias parameters $b_1$, $b_2$ and $b_{K^2}$. Indeed we expect results for third order bias parameters to be less accurate since we do not include $4^{\rm th}$ order ones in the bias expansion, and we did not perform thorough convergence tests to validate them.

%%%%%%%%%%%%%%%%%%%%%%%%%%%%%%%%%%%%%%%%%%%%%%%%%%%%%%%%%%%%%%%%%%%%%%%%%%%%%

\subsection{Expected dependence on the cutoff scale}
\label{sec:cutoff}

Due to our renormalization procedure that includes only renormalization of quadratic operators with respect to $\d_m$, as well as the fact that we stop at third order when constructing the bias expansion, we expect the bias parameters we measure with the method presented above to depend on the cutoff scale $\Lambda$. Specifically, the bias operators whose coefficients we measure here are constructed out of the sharp-$k$ filtered density field $\delta_{m,\Lambda}$, and thus explicitly refer to the scale $\Lambda$. On the other hand, the bias coefficients identified when measuring halo $N$-point functions in the large-scale limit formally refer to bias operators $[\O[\delta_m]]_0$, where $\delta_m$ is the unfiltered evolved field, which obey the \emph{renormalization conditions} given in \cite{assassi/etal} (see also section 2.10 of \cite{Desjacques:2016})
\be
\left\langle \left[\O[\delta_m]\right]_0 (\vk) \delta_m^{(1)}(\vk_1)\cdots \delta_m^{(1)}(\vk_n)\right\rangle
\stackrel{k_1,\ldots k_n\to 0}{\longrightarrow} \left\langle \O[\delta_m](\vk) \delta_m^{(1)}(\vk_1)\cdots \delta_m^{(1)}(\vk_n)\right\rangle_{\rm tree},
\label{eq:renormcond}
\ee
where the subscript ``tree'' indicates the tree-level perturbation theory result for the given correlator, and the superscript $^{(1)}$ indicates linear order. First, notice that our operators $\O[\delta_{m,\Lambda}]$ satisfy \refeq{renormcond} in the limit $\Lambda\to 0$. One can show that \cite{2020JCAP...11..008S}
\be
\O[\delta_{m,\Lambda}] \stackrel{\Lambda\to 0}{\longrightarrow} \left[\O[\delta_m]\right]_0 + \mathcal{\O}(\sigma_\Lambda^2),
\label{eq:Lambdadep}
\ee
where $\sigma_\Lambda^2 = \langle (\delta^{(1)}_{m,\Lambda})^2\rangle$ is the
variance of the linear, sharp-$k$ filtered density field. The $\Lambda$-dependent correction in \refeq{Lambdadep}
is not negligible in practice. For $n=1$, the renormalization
procedure that we apply to the quadratic operators indeed ensures \refeq{renormcond} (note that the right-hand side vanishes if $\O$ is a quadratic operator and $n=1$). 
We do not enforce the renormalization conditions at $n=2$ though, since
the non-vanishing right-hand side makes these conditions more difficult
to implement. Hence, we expect the operators used here to differ from the
large-scale renormalized operators by a correction $\propto \sigma_\Lambda^2$,
which correspondingly propagates to the bias parameters.
Higher-than-third-order bias terms that we do not include in our bias expansion also contribute a similar correction to the
measured bias parameters \cite{2020JCAP...11..008S}; what fraction of the
observed $\Lambda$-dependence is due to each of these contributions is
difficult to determine.

We thus expect the dependence of the bias parameters on the cutoff scale $\Lambda$ to be $b_\O(\Lambda) = b_\O + \tilde c_\O \sigma_\Lambda^2$.
In practice, $\sigma_\Lambda^2$ can be well approximated by a $\textrm{constant}\times\Lambda^2$ over the range of $\Lambda$ considered here.
Hence, we parametrize the dependence of the inferred bias parameters on $\Lambda$ as
\be
b_\O(\Lambda) = b_\O + c_\O \Lambda^2,
\label{eq:fitoflbda}
\ee
where $b_\O$ is the value of the bias parameter at $\Lambda \rightarrow 0$ that we quote as the forward modeling result. We expand further on this point in \refsec{bsq} where we use three cutoff values $\Lambda \in \{ 0.05, \, 0.075, \, 0.1\} \iMpch$ to perform a fit of the form of \refeq{fitoflbda} for each bias parameter, and obtain the desired value as the constant coefficient of the fit.

%%%%%%%%%%%%%%%%%%%%%%%%%%%%%%%%%%%%%%%%%%%%%%%%%%%%%%%%%%%%%%%%%%%%%%%%%%%%%
%%%%%%%%%%%%%%%%%%%%%%%%%%%%%%%%%%%%%%%%%%%%%%%%%%%%%%%%%%%%%%%%%%%%%%%%%%%%%
\section{Simulations and numerical data}
\label{sec:sims} 

In this section, we briefly describe the characteristics of our simulations and the procedure used to identify halos, determine their properties, and create the catalogs.

We use the same simulation set as the one referred to as ``L2400'' in \cite{Lazeyras:2017}, which is composed of two gravity-only simulations ran with the cosmological $N$-body code Gadget-2 \cite{Springel:2005}. We adopt a flat $\Lambda$CDM cosmology with $\Omega_m=0.27$, $h=0.7$, $\Omega_b h^2=0.023$, $n_s=0.95$ and $\mathcal{A}_s=2.2 \cdot 10^{-9}$. This is the same cosmology as the one used for the study of $b_1$ and $b_2$ from separate universe simulations in \cite{Lazeyras:2015, Lazeyras:2016}. The simulations have a box size $L= 2400 \, \Mpch$ on each side and $N= 1536^3$ matter tracer particles; this yields a particle mass resolution $m_p= 3\cdot10^{11} \, M_\odot/h$. The two simulations are for two different realizations of the random phases of the initial conditions field, which were generated with 2LPT at an initial redshift $z_{\rm ini}= 99$ using the code described in \cite{sirko:2005}. Notice that we use 2LPT to initialize the simulations from which we measure the \textit{observed} halo field, whereas we use the corresponding Zel'dovich field at the same redshift with the same random phases to construct the evolved matter density field and \textit{deterministic} bias relation, as explained in \refsec{biasexp}.

%%%%%%%%%%%%%%%%%%%%%%%%%%%%%%%%%%%%%%%%%%%%%%%%%%%%%%%%%%%%%%%%%%%%%%%%%%%%%

\subsection{Halo catalogs}
\label{sec:catalogs}

The dark matter halos were identified using the Amiga Halo Finder (AHF) code \cite{Gill:2004,Knollmann:2009} at $z=0, 0.5$ and $1$. AHF identifies halos with a spherical overdensity (SO) algorithm, and we set the value of the overdensity to 200 times the background density to define halos. The bound particles are then used to calculate canonical properties of halos like the density profile, rotation curve, mass, spin, and sphericity. In this paper, we consider only the objects identified as the main halos, and not their subhalos. 

In our results below, we will first study the bias parameters as a function of total halo mass alone (i.e., without considering assembly bias) in \refsec{bsq}.
In this part of the analysis, which only relies on the masses of the halos, we discard halos with less than 24 bound particles within $r_{200}$ (the radius corresponding to an overdensity of 200 with respect to the background density), which sets a minimum halo mass of  $\log M = 12.85$ ($\log$ is the base 10 logarithm, and masses are implicitly given in units of $h^{-1} M_\odot$ throughout). We divide the halo catalog at a given redshift in logarithmic mass bins of width $\Delta\log M = 0.2$, with the center of the lowest mass bin at $\log M = 12.95$.

To ensure sufficient signal-to-noise in our assembly bias results, we adopt a coarser mass binning before we further group the halos in bins of different properties. We restrict our assembly bias results to halos with at least 200 particles to ensure reasonable convergence of all of the halo properties we consider. The smallest halos hence have a mass of $\log M = 13.76$, and we divide the resulting halo catalogs into mass bins of equal size in logarithmic scale $\Delta\log M=0.4$ (with the first bin thus centered around $\log M = 13.96$). For each secondary property $p$ considered, we divide the subcatalogs at each mass in 4 quartiles, and we measure the bias parameters in each quartile to obtain the $b_\O(p | M)$ relation.

We will show the assembly bias dependence of $b_1$, $b_2$ and $\bkk$ on three halo properties: the concentration $c$, spin parameter $\lambda$, and sphericity $s$. The halo concentration is quantified using the usual Navarro-Frenk-White (NFW) \cite{Navarro:1996} concentration parameter $c$ measured as in \cite{Prada:2011}. More specifically, AHF computes the ratio between the maximum of the circular velocity $V_{\rm max}$ and $V_{200}$, the circular velocity at $r_{200}$. For the case of the NFW halo profile \cite{Navarro:1996}, this ratio is given by
\be
\frac{V_{\rm max}}{V_{200}}=\left(\frac{0.216 \, c}{f(c)}\right)^{1/2},
\label{eq:cprada}
\ee
where $f(c) = \ln (1+c)-{c}/(1+c)$. Computing $c$ from the circular velocity at two different radii is hence straightforward. However, we note that this way of inferring the concentration is not as robust as a proper fit of the halo density profile. For the halo spin, we use the spin parameter as defined in \cite{Bullock:2000}
\be
\lambda = \frac{|\v{J}|}{\sqrt{2}MVr_{200}},
\label{eq:spinparam}
\ee
where the angular momentum $\v{J}$, the mass $M$ and the circular velocity $V$ are evaluated at $r_{200}$.
Finally, following the works of e.g.~\cite{Faltenbacher:2009, Lazeyras:2016}, we also measure the bias as function of halo sphericity, defined as
\be
s=\frac{r_3}{r_1},
\label{eq:shape}
\ee
where $r_1 > r_2 > r_3$ are the axes of the moment-of-inertia tensor of the halo particles and $s=1$ thus corresponds to perfectly spherical halos.

Finally, we follow \cite{Wechsler:2005, Lazeyras:2016,Lazeyras:2020suj} and, for the concentration $c$ and halo spin $\lambda$, we define
\be
p'= \frac{{\rm ln}(p/\bar{p})}{\sigma({\rm ln}p)},
\label{eq:prime}
\ee
where $p$ is the mean value of the property in a given quartile and mass bin, $\bar{p}$ is the mean value over all quartiles in a given mass bin, and $\sigma$ is the square root of the variance of ${\rm ln}p$ at fixed mass across all quartiles. We then plot the relations $b_1(p' | M)$. This can be done since these two properties are known to be approximately lognormal distributed at fixed mass \cite{Bullock:2000,Bullock:1999he}, and allows to remove most of the mass dependence of the values of these properties. Since this is not the case for halo sphericity, we do not adopt the same procedure for this property, and plot $b_1$ at fixed mass as a function of the mean sphericity $s$ in each quartile.  

%%%%%%%%%%%%%%%%%%%%%%%%%%%%%%%%%%%%%%%%%%%%%%%%%%%%%%%%%%%%%%%%%%%%%%%%%%%%%
%%%%%%%%%%%%%%%%%%%%%%%%%%%%%%%%%%%%%%%%%%%%%%%%%%%%%%%%%%%%%%%%%%%%%%%%%%%%%

\section{Results and discussion}
\label{sec:results}

In this section, we begin by presenting the dependence of the bias parameters on the cutoff scale $\Lambda$, focusing first on results for the full mass sample (i.e without binning in a secondary property, \refsec{bsq}). We then show results for assembly bias in \refsec{assembly}, focusing on subsamples for the three halo properties described in the previous section. Finally, we investigate the impact of assembly bias on the relations $b_2(b_1)$ and $\bkk(b_1)$ in \refsec{impact}.

Notice that the number of objects in some samples can become quite low, especially at high mass when we bin in secondary properties. The correlation between these samples and the dark matter field, from which we essentially measure the bias parameters, can in these cases become noise dominated and affect the inference of the bias parameters. We hence decide which samples are robust based on the following criterion (which is effectively the same as in our companion paper \cite{Barreira:2021ukk}). For all binned samples, we evaluate the phase-correlation coefficient between halos and matter, $r(k) =P_{hm}(k)/\sqrt{P_{mm}(k)P_{hh}(k)}$, where $P_{mm}$, $P_{hh}$ and $P_{hm}$ are the matter, halo and corresponding cross power spectra at a given redshift, respectively. At low $k$, the matter power spectrum is much larger than the halo shot noise and so $r(k)$ is close to unity, i.e., the halo and matter fields are strongly correlated on large scales.  Towards higher $k$, the shot noise contribution remains roughly constant, but since the matter power spectrum becomes smaller, the correlation coefficient drops. We will thus only show results for halo samples that satisfy $r(k) \geq 0.5$ at $k = \Lambda$. This roughly ensures that our halo samples are still not stochastic-dominated at the scale of the cutoff $\Lambda$ that we use, and thus, that the EFT likelihood formalism can utilize the correlation between the halo and matter fields to infer the bias parameters. 

\subsection{Dependence on the cutoff scale $\Lambda$}
\label{sec:bsq}

\begin{figure}[t]
\centering
\includegraphics[scale=0.5]{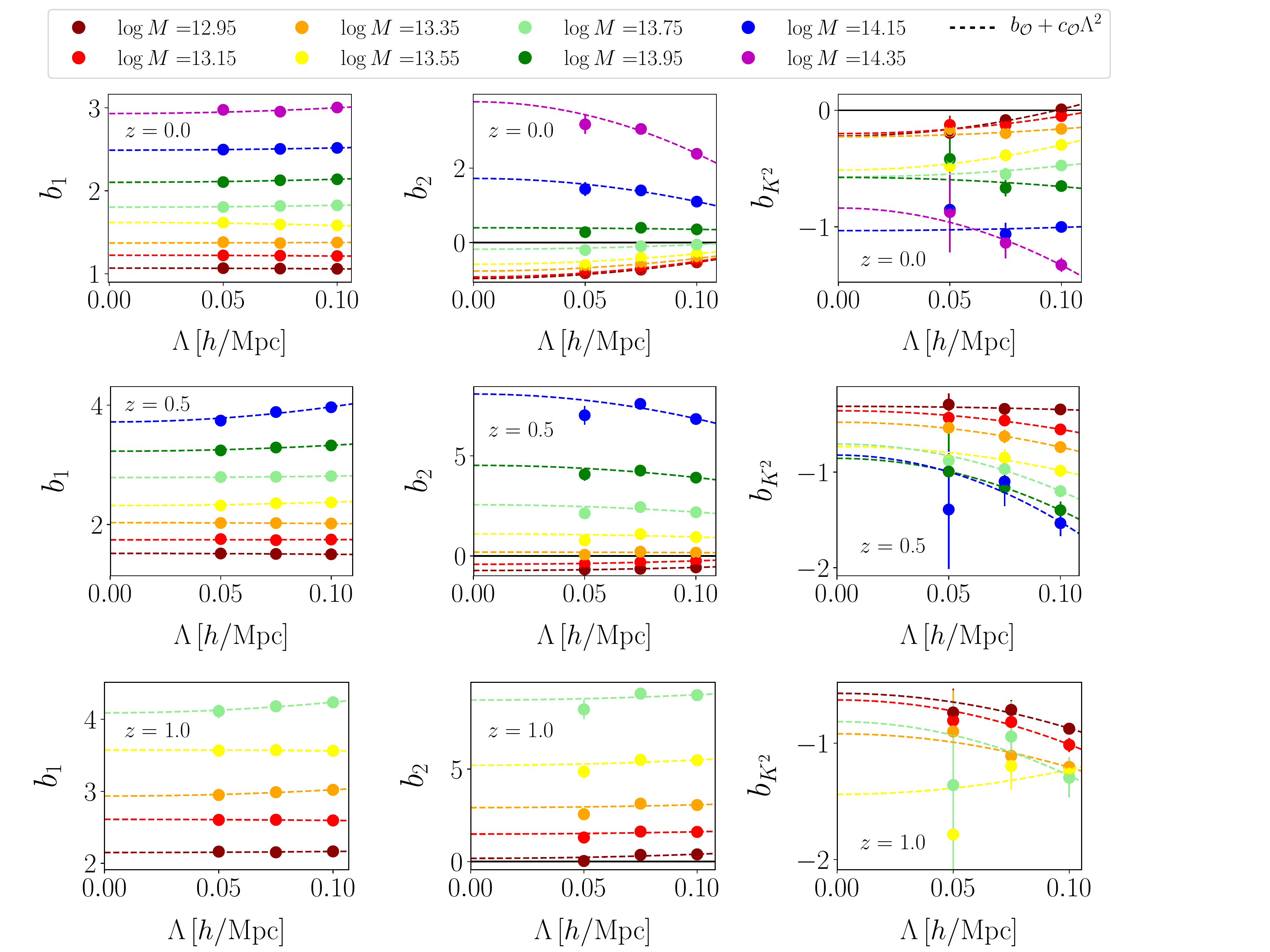}
\caption{Dependence of $b_1$ (left column), $b_2$ (middle column) and $b_{K^2}$ (right column) on the cutoff scale $\Lambda$ at fixed mass and at $z=0, \, 0.5, \, 1$ (top, middle, and bottom row respectively). The points present measurements obtained from forward modeling for different values of $\Lambda$, and the dashed lines show the respective fits of the form of \refeq{fitoflbda}. The different colors are for different mass bins, as labeled. We see that the expected relation is in general respected, which allows us to extrapolate the value of the parameters at $\Lambda \rightarrow 0$.}
\label{fig:bofLbda}
\end{figure} 

We first focus on the dependence of the bias parameters on the cutoff scale $\Lambda$ at fixed halo mass. As explained before, in this section, we use halo catalogs in mass bins of width $\log M = 0.2$, and with halos containing down to 24 particles. \refFig{bofLbda} presents this dependence for all mass bins satisfying $r(k=0.1 \iMpch)>0.5$ (the colors indicate the mass bins, as labeled) at $z=0$ (top row), $0.5$ (middle row), and $1$ (bottom row). The left, middle and right panels show the results for $b_1$, $b_2$ and $\bkk$, respectively. The points present measurements obtained from forward modeling at fixed $\Lambda$ values, while the dashed lines show the corresponding fit of the form of \refeq{fitoflbda}. We see that the expected dependence is in general well respected, which allows us to obtain the value of the parameters by extrapolating to $\Lambda = 0$. The fit is typically dominated by the $\Lambda=0.1 \, \iMpch$ result, which is the most statistically precise, but which leads to extrapolation uncertainties when the $\Lambda$-dependence is strong (this is mostly visible for $b_2$ at high mass). In the case of $b_1$, the $\Lambda$-dependence is less marked. This indicates that the third-order bias expansion we use in this work, as well as our renormalization procedure in which we renormalize second order operators only with respect to $\d_m$,  are sufficient to remove most of the $\Lambda$ dependence of the linear bias. A higher-order analysis and improved renormalization procedure should reduce the $\Lambda$-dependence of the quadratic bias coefficients to a similar level as seen for $b_1$ here.

\begin{figure}
\centering
\includegraphics[scale=0.43]{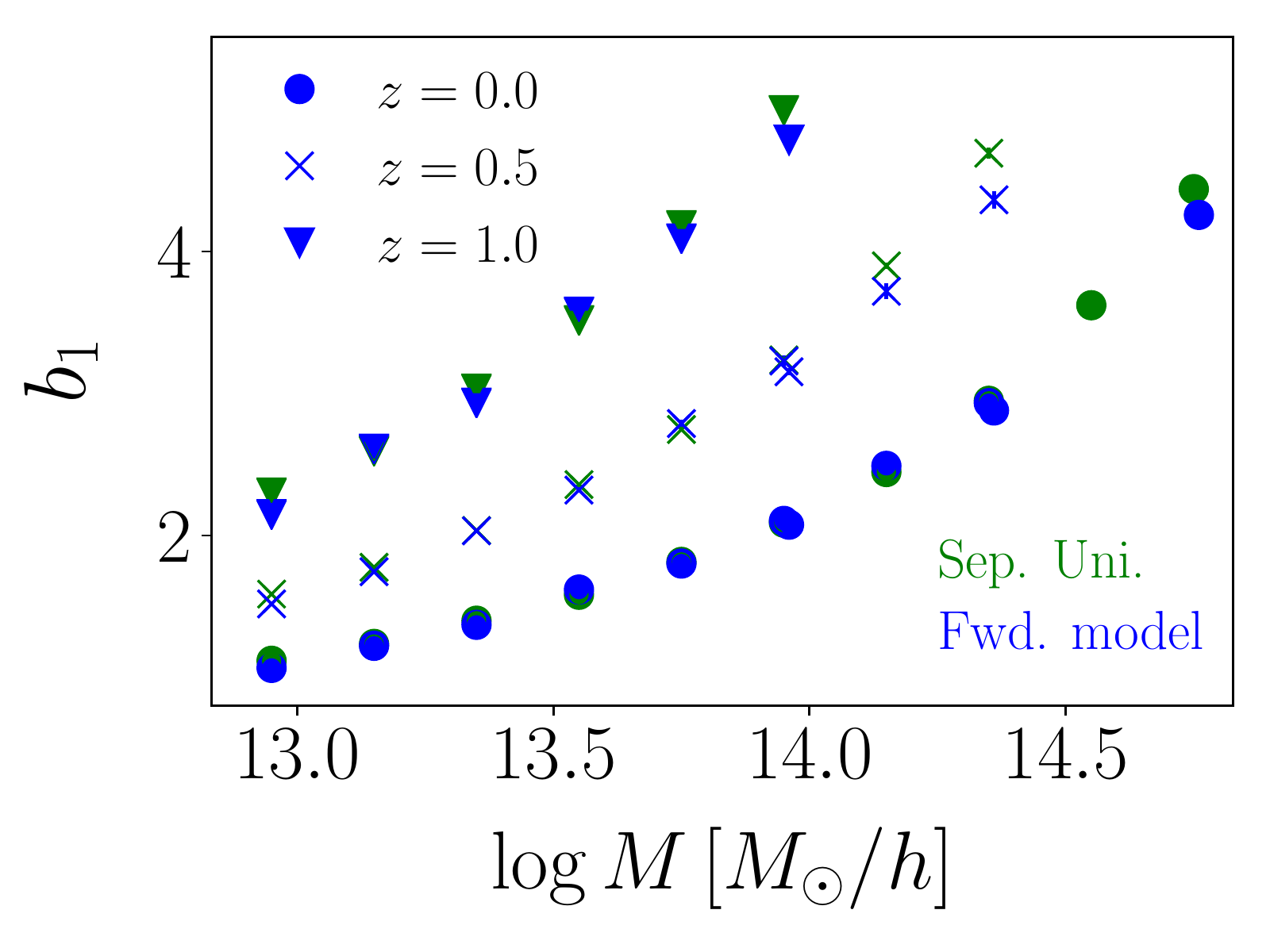}
\includegraphics[scale=0.43]{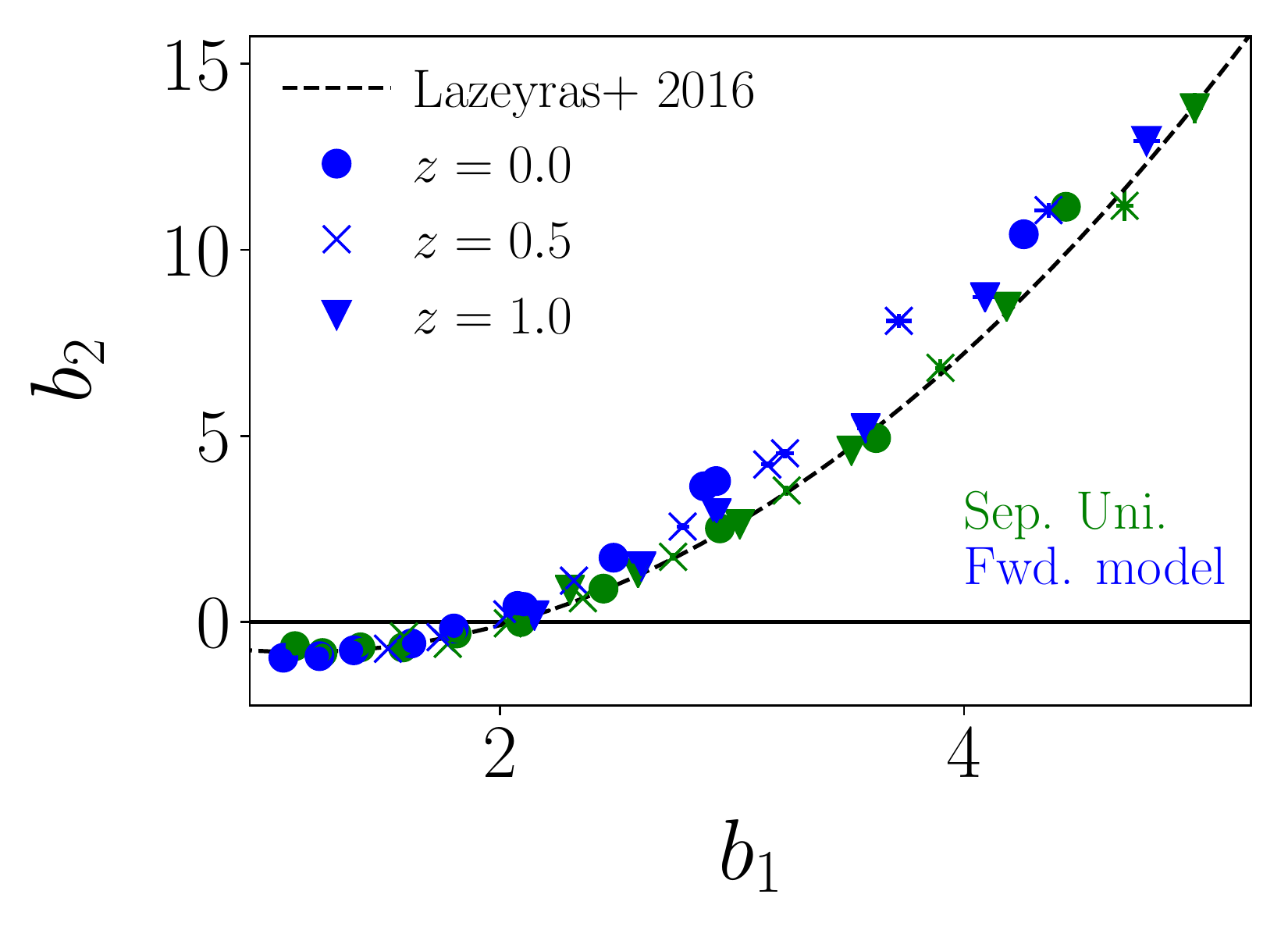}
\includegraphics[scale=0.45]{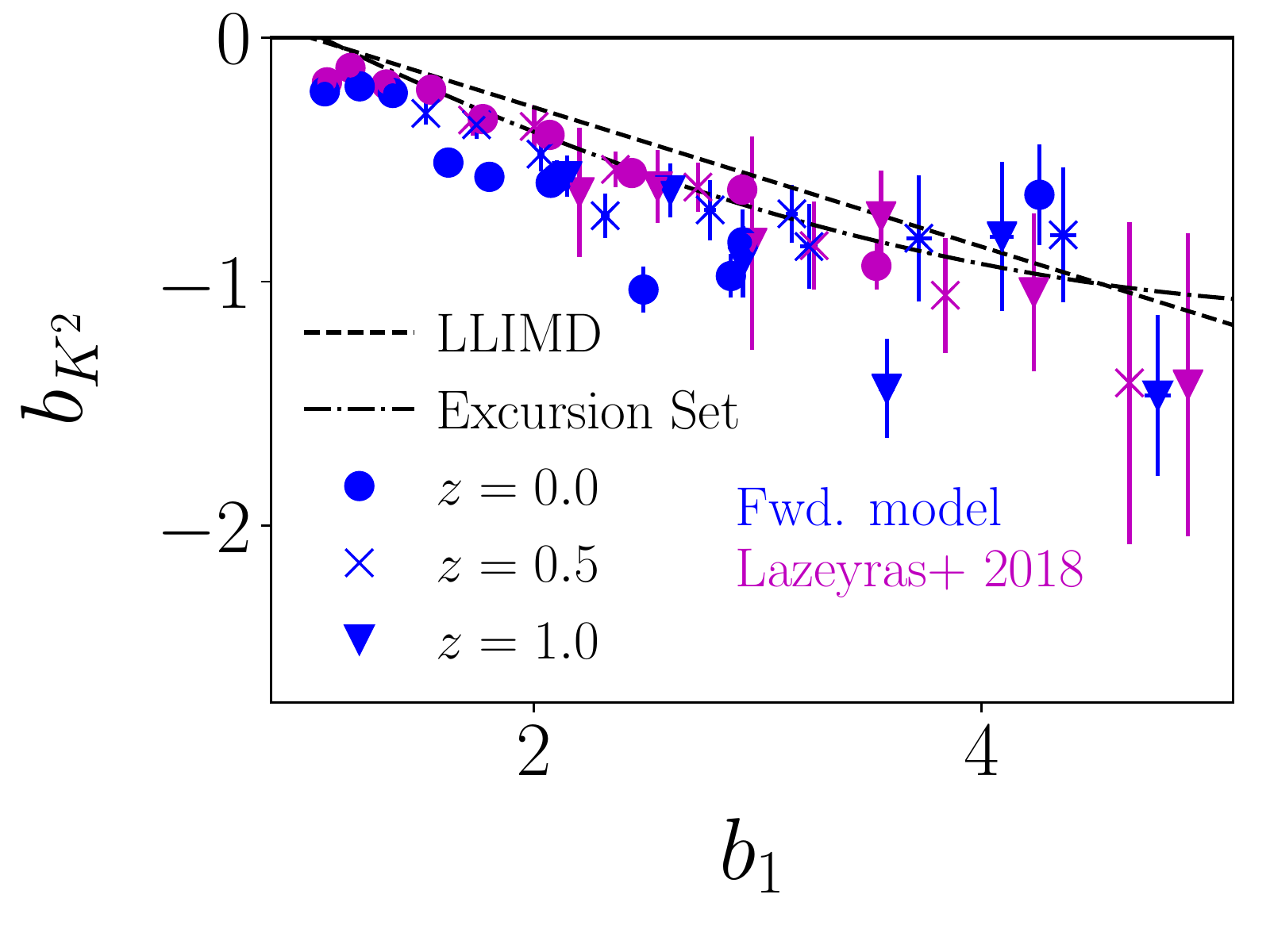}
\caption{\textbf{Top left:} Linear bias parameter $b_1$ as a function of mass. The blue symbols present our results from the forward model at $\Lambda \rightarrow 0$, and we compare them to the green points obtained with separate universe simulations in \cite{Lazeyras:2015}. The form of the markers indicates the redshift. As can be seen, we find excellent agreement between the two methods. \textbf{Top right:} Second order bias parameter $b_2$ as a function of $b_1$. The points correspond to the same halo samples shown in the top left panel; the markers and color coding are also the same. We compare to separate universe results, and the best fit obtained from those (dashed line)  \cite{Lazeyras:2015}. The forward modeling results agree broadly with the separate universe ones. We attribute the small offset at $b_1 \gtrsim 2.5$ to higher-order bias parameters that we did not include, as well as to our renormalization procedure. \textbf{Bottom:} Same as the top right panel but for the tidal bias parameter $\bkk$. This time we compare our results to those presented in \cite{Lazeyras:2017} (purple markers), and find good agreement. When comparing to the LLIMD prediction (dashed, \refeq{LLIMD}), the already known small negative offset is visible. We also compare our $\bkk$ results to \refeq{bK2ES}, which is a prediction from the excursion set model with a barrier inspired from triaxial collapse \cite{Sheth:2012fc}, that performs slightly better than the LLIMD prediction.}
\label{fig:bsqofb1}
\end{figure} 

We now turn to \reffig{bsqofb1} which presents results for $b_1$ as a function of halo mass, as well as $b_2$ and $\bkk$ as a function of $b_1$. The forward modeling results (blue symbols) plotted in this figure are the ones extrapolated at $\Lambda \rightarrow 0$. However, since the inferred error on the parameters of the fit turns out to be rather unstable\footnote{We use the built-in python function scipy.optimize.curve\_fit to perform the fits in \reffig{bofLbda}.}, the errorbars shown in \reffig{bsqofb1} are simply the ones of the corresponding point at $\Lambda = 0.075 \iMpch$, which are the marginalized errors returned by the {\sc minuit} minimizer. Indeed, we found these to be more stable, and to always be slightly larger than the ones returned by the fitting routine. The results for $b_1$, presented on the top left panel, are in very good agreement with those from separate universe simulations (green markers) \cite{Lazeyras:2015}. This panel provides the correspondence between halo mass and linear bias which can be useful for a detailed inspection of the upcoming figures, which will all be in terms of $b_1$. 

The results for $b_2(b_1)$, presented in the top right panel of \reffig{bsqofb1}, are also in good agreement with those from separate universe simulations \cite{Lazeyras:2015} (green markers and dashed line). We do note, however, a small trend for higher $b_2$ values compared to the separate universe result for $b_1 \gtrsim 2.5$. This is likely due to the fact that we neglect higher-than-third order bias parameters (which can become important for these more biased objects), and that we do not perform the complete renormalization of operators necessary to compare the forward-modeling bias parameters with those obtained in the large-scale limit, such as in the separate universe technique (\refsec{cutoff}). Further, as we noted above, the point at $\Lambda=0.1 \iMpch$ dominates the fit because it uses a larger number of modes, but it is also the one that is most impacted by these higher-order effects; this can affect the extrapolation to $\Lambda \to 0$ and thus help explain some of the offset in the top right panel of \reffig{bsqofb1}. In future work, we plan to include higher-order biasing and refine the renormalization procedure, which would clarify the origin of these small differences.

The results for $\bkk$ as a function of $b_1$ (bottom panel of \reffig{bsqofb1}) are also in broad agreement with previous results from \cite{Lazeyras:2017} (purple markers). We further compare our results for $\bkk$ with the Lagrangian local-in-matter-density (LLIMD) prediction given by 
\be
\bkk^{\rm LLIMD} = -\frac27 (b_1 -1),
\label{eq:LLIMD}
\ee
which assumes a Lagrangian tidal bias parameter $\bkk^L =0$, i.e., it neglects the impact of the tidal field in the initial conditions and considers only the tidal field generated by gravitational evolution (see e.g.~sections 2.2-2.4 of \cite{Desjacques:2016}). We observe in our results the same slightly negative offset already pointed out in \cite{Sheth:2012fc,Saito:2014qha,Lazeyras:2017,Abidi:2018,Eggemeier:2020umu,Eggemeier:2021cam}, which is unsurprising given the simplified assumption behind the LLIMD result. We make a comparison also with a more elaborate model developed in \cite{Sheth:2012fc}, coming from the excursion set formalism (\cite{Bond:1990}) with correlated steps and a barrier inspired from triaxial collapse. This model allows for $\bkk^L \neq 0$, and the prediction for $\bkk$ can be represented by the following simple quadratic fit 
\be
\bkk^{\rm ES} = 0.524 - 0.547 b_1 + 0.046 b_1^2.
\label{eq:bK2ES}
\ee
We see that this prediction seems to perform slightly better than the LLIMD prediction, as was already noted in \cite{Eggemeier:2020umu}. Given the scatter of our measurements, we refrain from producing our own fit and defer this to future work when the $\bkk(b_1)$ relation will be more tightly constrained (see \cite{Eggemeier:2021cam} though).

%%%%%%%%%%%%%%%%%%%%%%%%%%%%%%%%%%%%%%%%%%%%%%%%%%%%%%%%%%%%%%%%%%%%%%%%%%%%%

\subsection{Assembly bias}
\label{sec:assembly}

\begin{figure}
\centering
\includegraphics[scale=0.52]{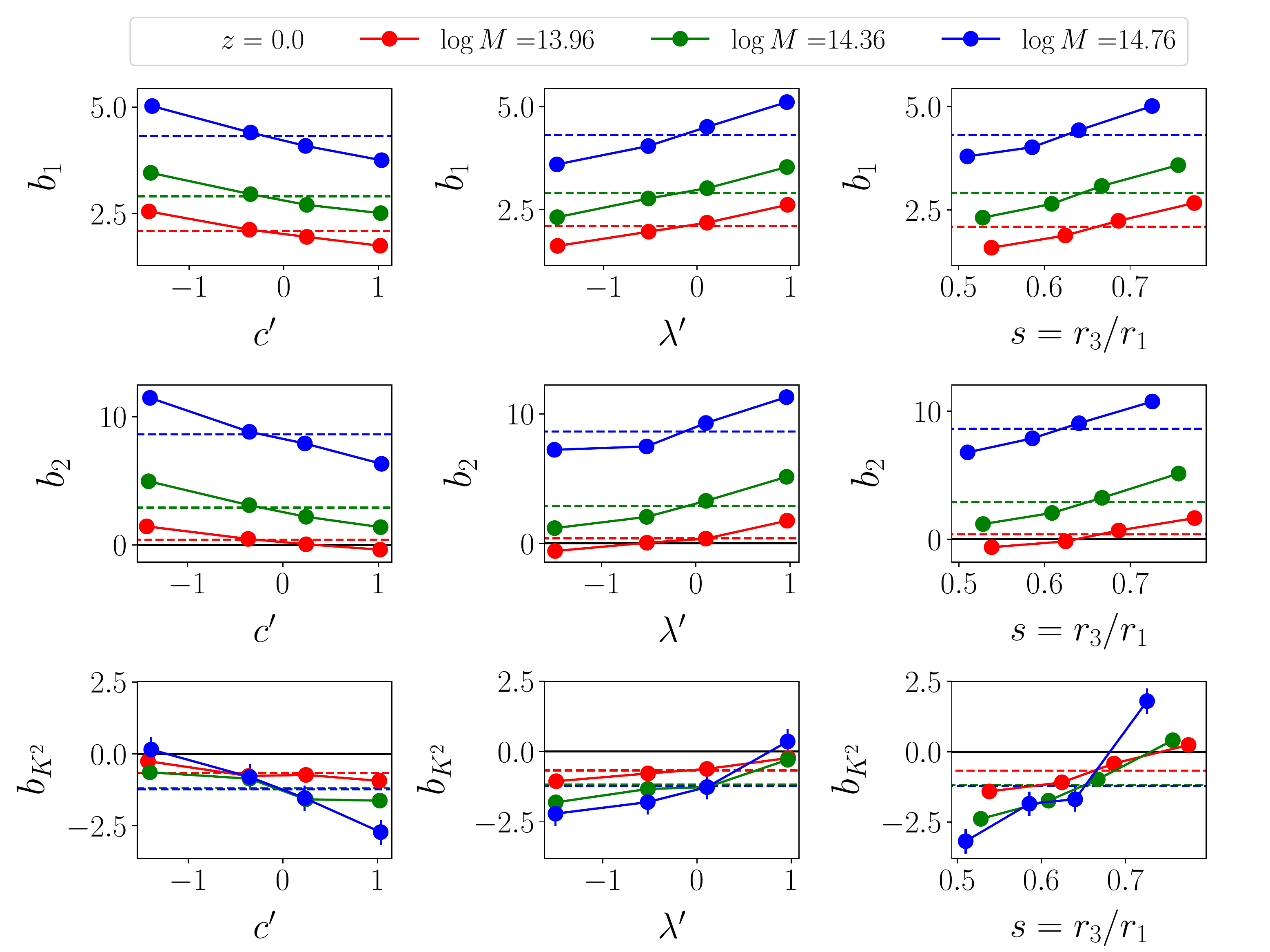}
\caption{Assembly bias in $b_1$ (top row), $b_2$ (middle row), and $\bkk$ (bottom row) as a function of concentration (left), spin (middle), and sphericity (right) for various mass bins (color coded) at $z=0$. The points are the forward model results for $\Lambda=0.075$ $\iMpch$. The dashed lines indicated the mean value of the bias in each mass bin. On each panel, halos that are more concentrated, have higher spin, or are more spherical are on the right. We get a clear detection of assembly bias in all three parameters. In the case of $b_1$ and $b_2$, the trends are in qualitative agreement with previous results. The assembly bias signal in $\bkk$ is new to this work, and we find it to be consistently opposite to the one in the $b_1$ parameter (e.g.~halos that are more positively biased in $b_1$ are less negatively biased in $\bkk$).}
\label{fig:bK2ofp}
\end{figure}

In this section we present our results for assembly bias in the three bias parameters considered in this work. All the results presented next are for fixed $\Lambda = 0.075 \, \iMpch$, i.e.~we do not perform the quadratic fit in $\Lambda$ to extrapolate values at $\Lambda\rightarrow 0$, since we are mostly interested in the assembly bias trends, rather than the precise value of bias parameters in a particular bin. This increases the overall number of samples satisfying our $r(k = \Lambda) \geq 0.5$ criterion while keeping a good signal-to-noise ratio. Furthermore, we can see from \reffigs{bofLbda}{bsqofb1} that this should not add any strong systematics to our results with respect to those at $\Lambda \rightarrow 0$. We recall that we use wider mass bins ($\Delta \log M = 0.4$) for the results in this section since we further bin in a secondary halo property. Note also that although we refer to low and high masses in our discussion next, one should keep in mind that all our results are technically for massive halos (i.e., group- or cluster-sized halos), since we here consider only halos with more than 200 particles. Furthermore, as explained in \refsec{sims}, we used a spherical overdensity algorithm to identify halos. This most likely has an impact on our findings \cite{Chue:2018}, but we do not investigate how they change if we used, e.g., a friends-of-friends (FoF) algorithm.

\refFig{bK2ofp} presents results for assembly bias in $b_1$ (top row), $b_2$ (middle row), and $\bkk$ (bottom row) as a function of concentration, spin, and sphericity, and for different halo mass bins at $z=0$, as labeled. The dashed lines show the value of the bias parameter in the corresponding total mass bin. On each panel, from left to right, the halos become more concentrated, have higher spin, and become more spherical. Focusing first on $b_1$ and $b_2$ for which previous results exist, we get a very clean detection of assembly bias. These results are also in agreement with previous ones from the literature (e.g.~\cite{Lazeyras:2016}). In particular, we find that the very massive halos considered in this work become less biased in $b_1$ and $b_2$ as concentration increases, become more biased as spin increases, and that more spherical halos ($s \rightarrow 1$) are also more biased. These results provide a further check of the robustness of the forward model results, and will help us analyze the results for the impact of assembly bias on the $b_2(b_1)$ and $\bkk(b_1)$ relations in the next subsection.   

Turning now to the bottom row of \reffig{bK2ofp}, we see that we obtain a clear detection of assembly bias in $\bkk$ as well, and that the effect seems more important at higher masses for all three properties. However, the trends observed in this parameter are \textit{consistently opposite to those of $b_1$;} this is one of the main findings of this paper. In particular, halos with a higher concentration are \textit{more negatively biased} in $\bkk$, which is the opposite to the trend in $b_1$ in the sense that the halos are {\it less positively biased}. Further, halos with a higher spin have $\bkk$ closer to zero, but $b_1$ becomes larger. Finally, more spherical halos are \textit{less negatively biased} in the case of $\bkk$, which is again opposite to what is found for $b_1$. 

\subsection{Impact of assembly bias on the $b_2(b_1)$ and $\bkk(b_1)$ relations}
\label{sec:impact}

\begin{figure}[t]
\centering
\includegraphics[scale=0.45]{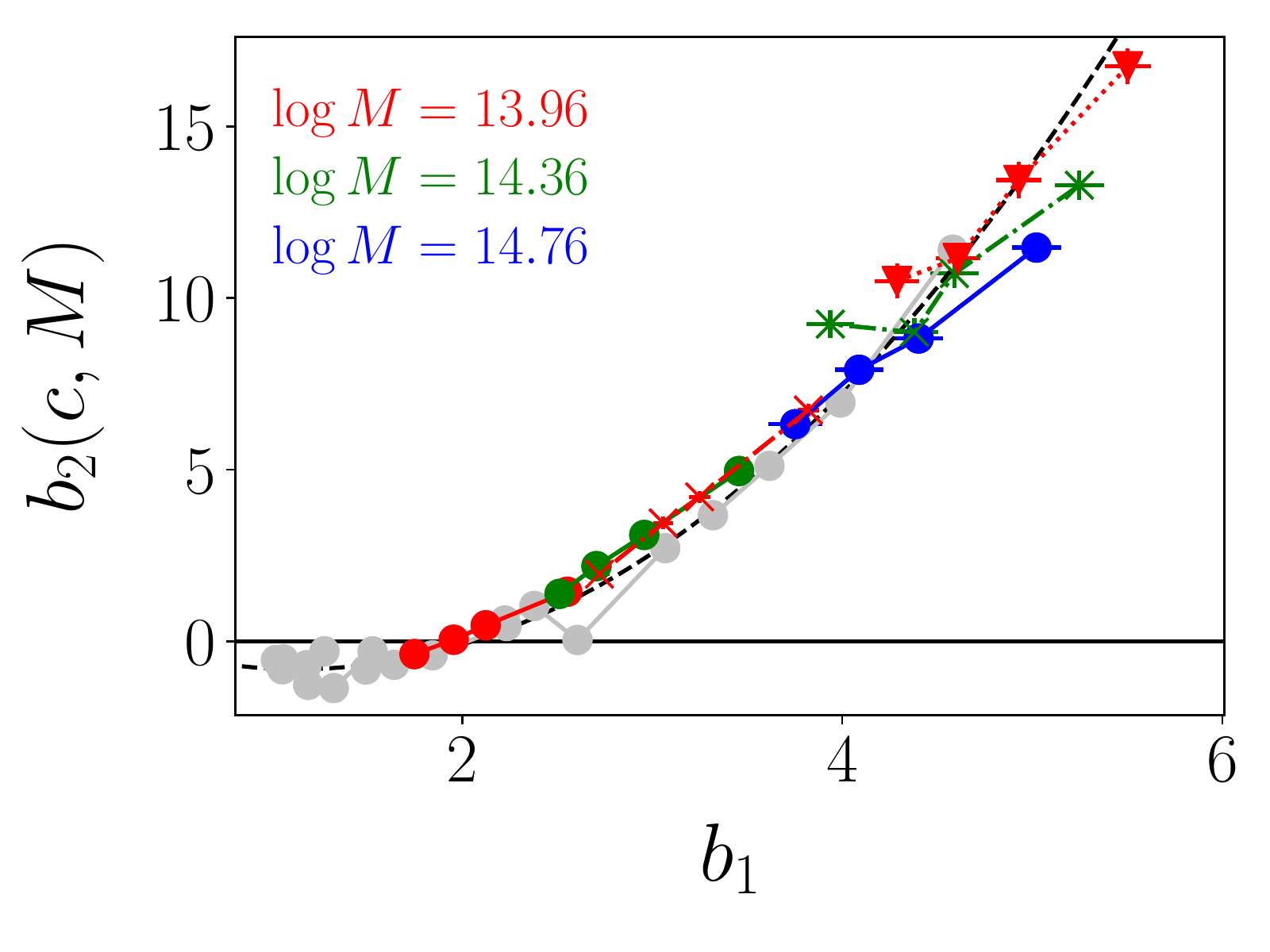}
\includegraphics[scale=0.45]{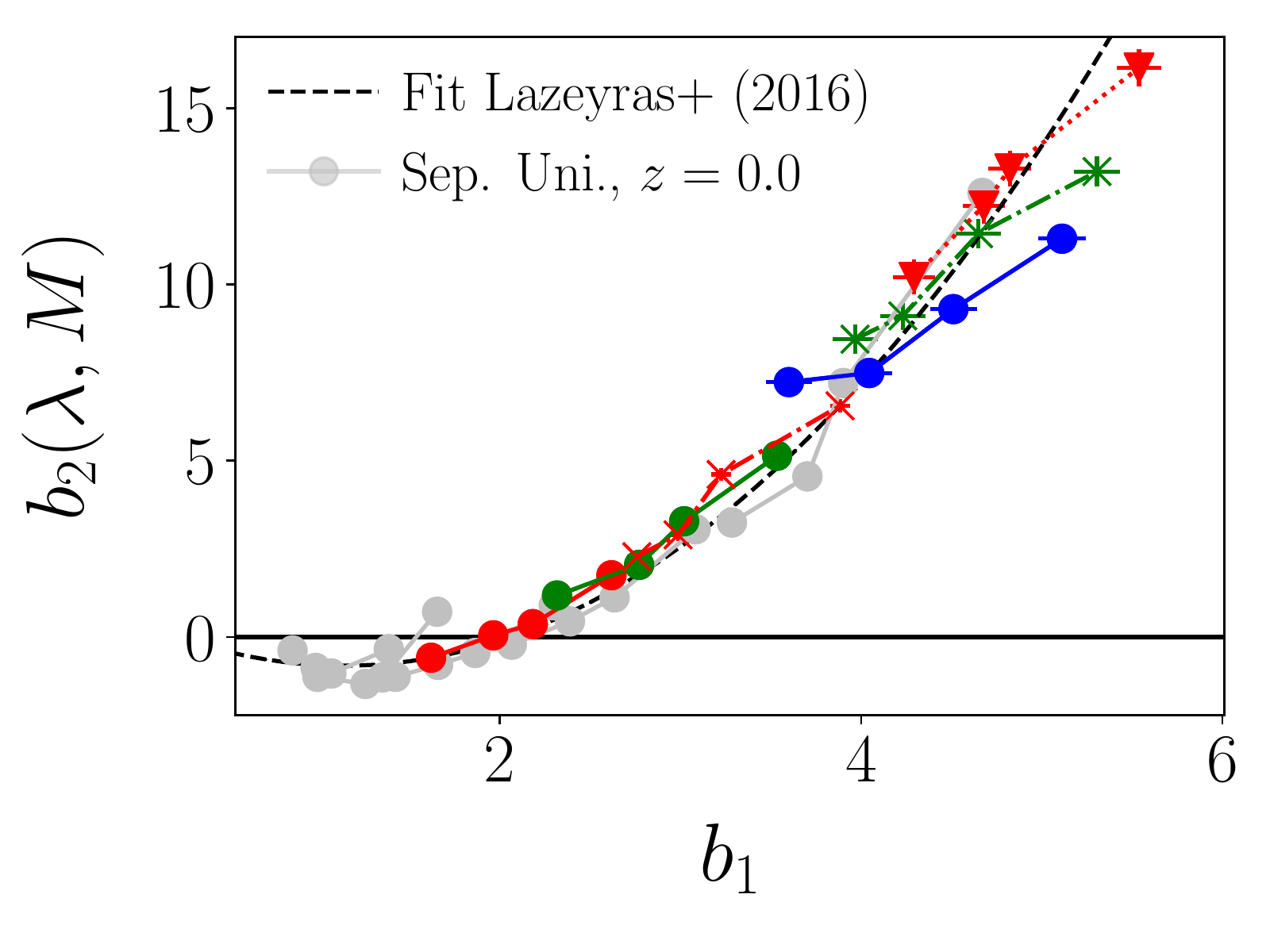}
\includegraphics[scale=0.45]{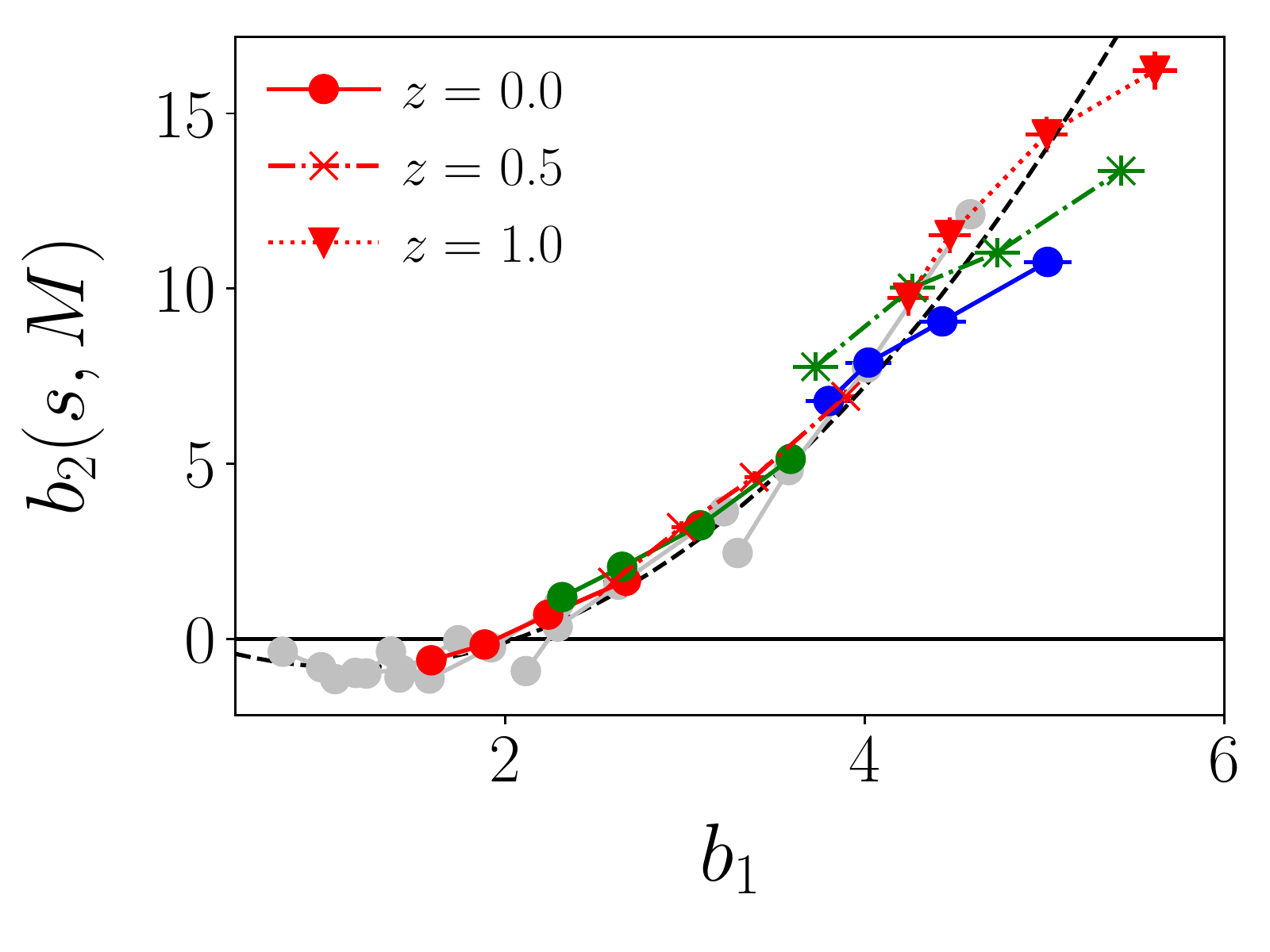}
\caption{The impact of assembly bias on $b_{2}$ as a function of $b_1$ at different mass (color coding) and redshift (indicated by the symbols and linestyles). The top left, top right and bottom panels show, respectively, the results for concentration, spin and sphericity as the secondary halo property (four same colored symbols). The results are for $\Lambda=0.075$ $\iMpch$. The grey symbols show the assembly bias measurement of \cite{Lazeyras:2016} using separate universe simulations, which agree well with the forward model result. The dashed line shows the fitting function for $b_2(b_1)$ presented in \cite{Lazeyras:2015}, which remains broadly respected even after assembly bias is taken into account.}
\label{fig:b2ofb1ofp}
\end{figure} 

\begin{figure}[t]
\centering
\includegraphics[scale=0.45]{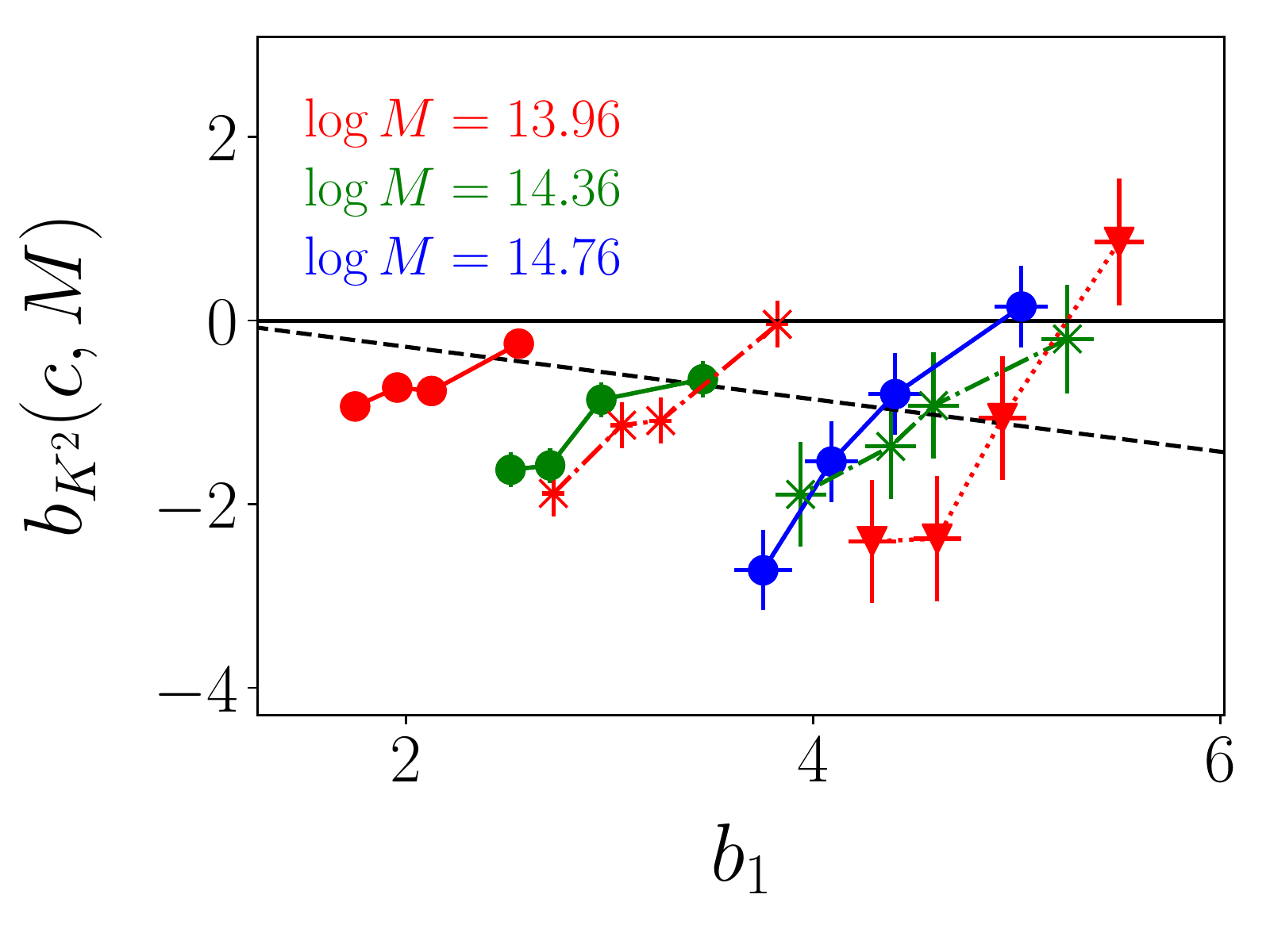}
\includegraphics[scale=0.45]{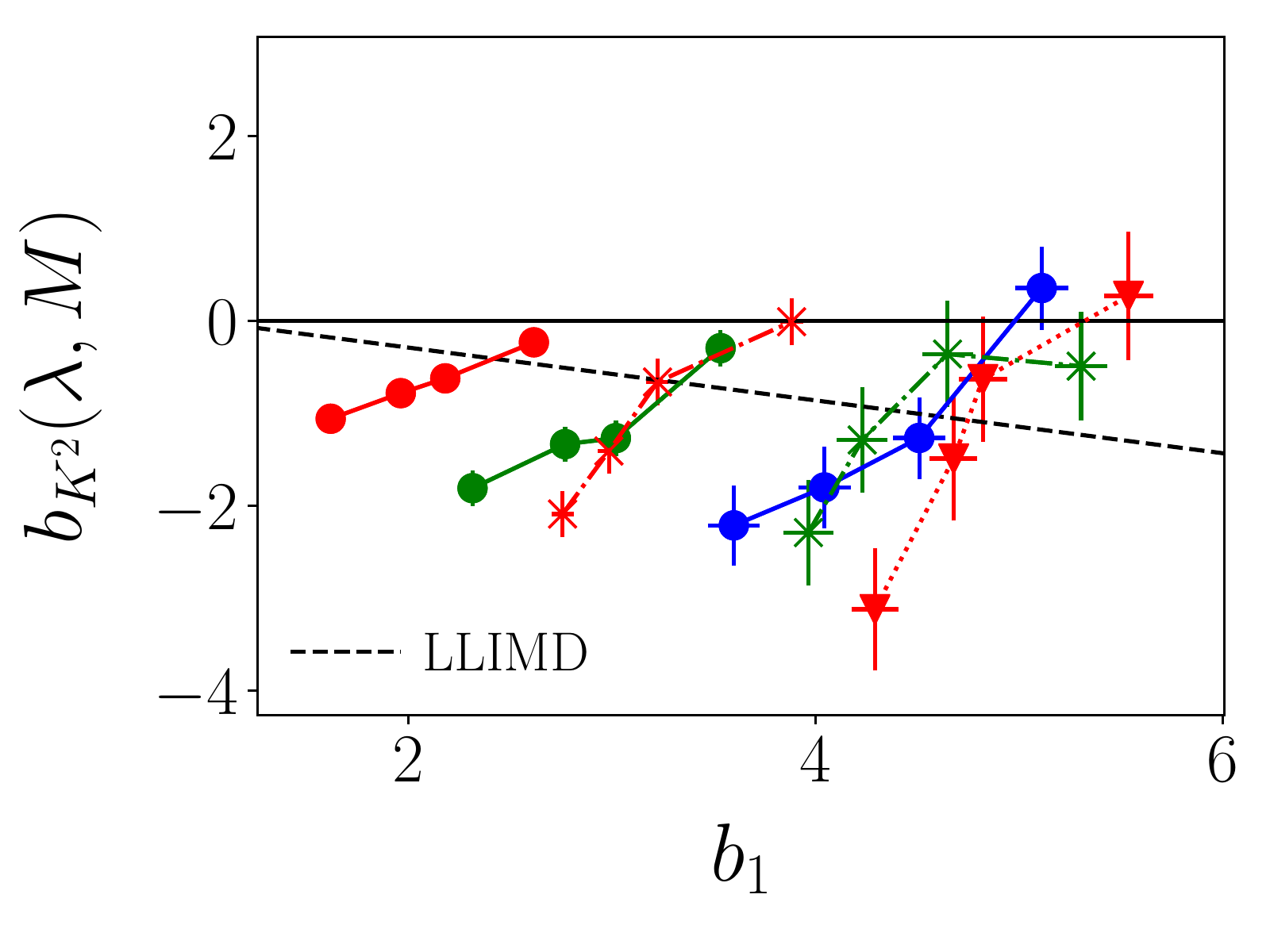}
\includegraphics[scale=0.45]{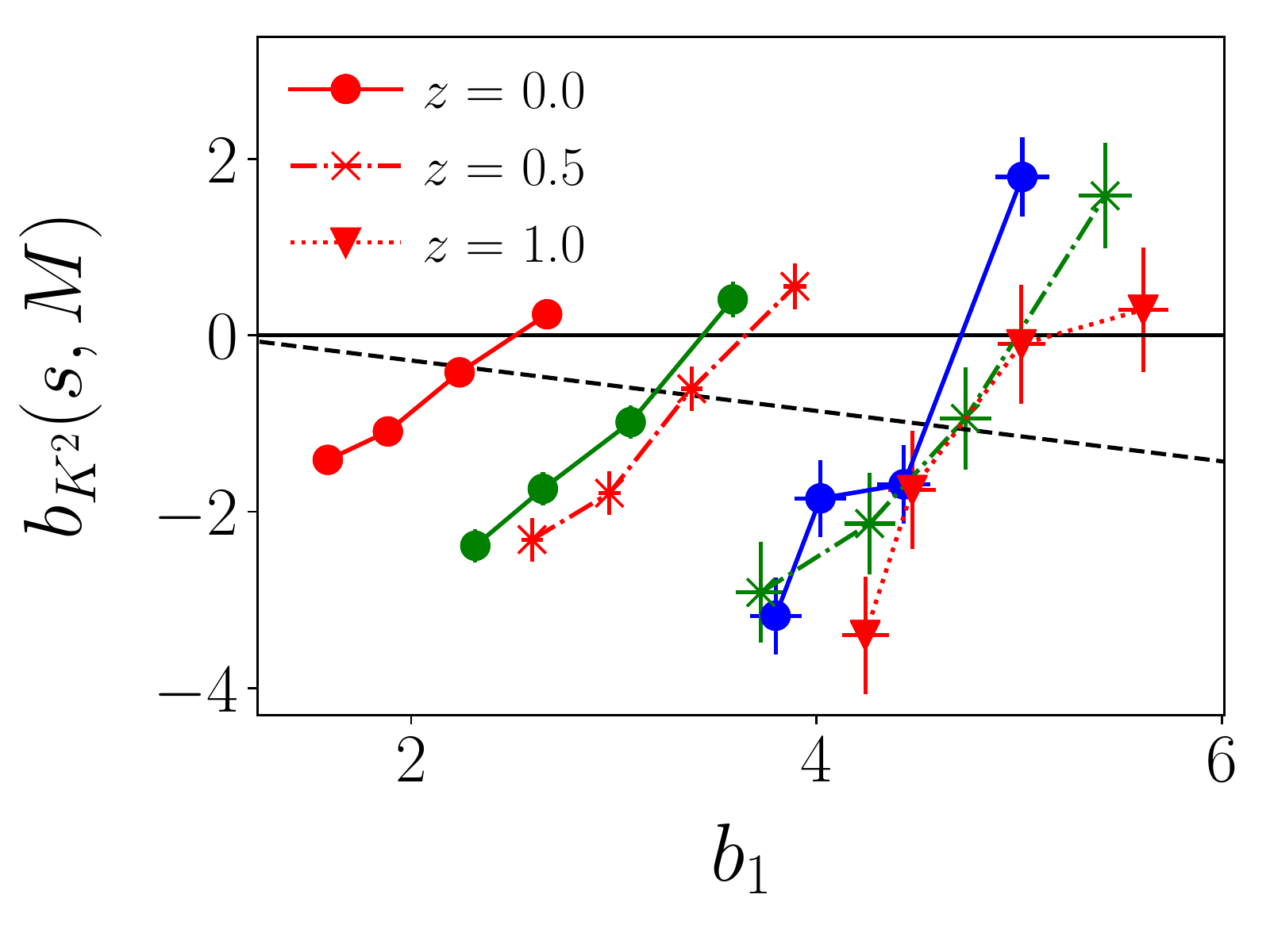}
\caption{Same as \reffig{b2ofb1ofp} but for the relation between $b_{K^2}$ and $b_1$. We show the LLIMD prediction (dashed line) as reference to indicate roughly where the mean bias relation lies. Here, contrary to the $b_2(b_1)$ case, assembly bias markedly modifies the $\bkk(b_1)$ relation, with the variation induced by the secondary properties at fixed halo mass being roughly orthogonal to the direction of the LLIMD relation.}
\label{fig:bK2ofb1ofp}
\end{figure}

We now investigate the impact of assembly bias on the $b_2(b_1)$ and $\bkk(b_1)$ relations. The results are presented in \reffigs{b2ofb1ofp}{bK2ofb1ofp}. Each panel of these figures presents these relations when binning in mass (color coding) as well as when binning in a secondary property (four markers of the same color). The different symbols and linestyles show results at different redshift. In \reffig{b2ofb1ofp} we use the fit from the separate universe simulations of \cite{Lazeyras:2015} as reference for the $b_2(b_1)$ relation. Also for comparison, the grey points show the assembly bias measurement from \cite{Lazeyras:2016}, obtained with separate universe simulations at redshift zero. We take the LLIMD relation as reference for $\bkk(b_1)$ in \reffig{bK2ofb1ofp} to guide the eye.

Firstly, in \reffig{b2ofb1ofp}, we see that our results from forward modeling agree quite well with those from separate universe simulations for all secondary properties, especially at lower masses. Perhaps the most noteworthy feature is the fact that, although assembly bias impacts the values of $b_2$ strongly (see middle row of \reffig{bK2ofp}), it only impacts the $b_2(b_1)$ relation weakly. The only discernible differences are seen for the most massive halos, where the relation becomes slightly shallower at all redshift and for all properties. This weak dependence of the $b_2(b_1)$ relation on assembly bias is in strong contrast with the results for $\bkk(b_1)$ in \reffig{bK2ofb1ofp}, where we see that, at fixed mass and for all redshift, the relation varies almost \textit{orthogonally} to the LLIMD prediction as the secondary halo properties vary. The surprisingly strong difference between the impact of assembly bias on the $b_2(b_1)$ and $\bkk(b_1)$ relations can be understood by inspecting  the dependence of the three bias parameters on $\{c, \, \lambda, \, s\}$ in \reffig{bK2ofp}. For example, focusing on the intermediate mass bin (green color), increasing the concentration makes  (i) $b_1$ less positive ; (ii) $b_2$ less positive and (iii) $\bkk$ more negative. Since $b_2(b_1)$ is monotonically increasing for $b_1 \geq 1.5$, if assembly bias shifts $b_2$ and $b_1$ in the same direction and by approximately the same amount, then the relation is preserved, as it seems to be the case (although to a lesser degree for the highest mass bin). On the other hand, for the same mass bin, increasing the concentration shifts $b_1$ and $\bkk$ in opposite directions, i.e., $\bkk$ goes down and $b_1$ goes left in the $\bkk - b_1$ plane, which results in the observed orthogonality. A similar reasoning holds for the other mass bins and other secondary properties. 

%%%%%%%%%%%%%%%%%%%%%%%%%%%%%%%%%%%%%%%%%%%%%%%%%%%%%%%%%%%%%%%%%%%%%%%%%%%%%
%%%%%%%%%%%%%%%%%%%%%%%%%%%%%%%%%%%%%%%%%%%%%%%%%%%%%%%%%%%%%%%%%%%%%%%%%%%%%

\section{Conclusion}
\label{sec:concl} 

In this paper, we have used the recently developed forward modeling formalism for galaxy clustering combined with the likelihood of LSS derived from EFT to study assembly bias in the three first bias parameters of dark matter halos, $b_1$, $b_2$ and $\bkk$. Assembly bias in $b_1$, $b_2$ has been studied before (and our results agree with past works), but this is the first time that these investigations are carried out also for the tidal bias $\bkk$. We focused on three secondary halo properties (halo concentration, spin, and sphericity), and we have investigated in particular the impact of assembly bias on the $b_2(b_1)$ and $\bkk(b_1)$ relations. Our main findings are:
\begin{itemize}

\item Our results for $b_1(M)$, $b_2(b_1)$ and $\bkk(b_1)$ binned in mass only (i.e., averaging over assembly bias) agree well with previous works (\reffig{bsqofb1}), which supports the validity of the forward modeling approach to study halo bias.

\item We detect assembly bias in all three parameters with high significance. One of our main findings is that assembly bias impacts $\bkk$ in a way that is consistently opposite to $b_1$ (for example, more concentrated halos have less positive $b_1$, but more negative $\bkk$; see \reffig{bK2ofp}).

\item Assembly bias does not impact the $b_2(b_1)$ relation significantly as it affects both $b_1$ and $b_2$ in a similar manner. On the other hand, the opposite trends observed in $\bkk$ strongly modify the $\bkk(b_1)$ relation, which at fixed halo mass, varies approximately orthogonally to the direction of the full mass sample (i.e. when binning in mass only; see \reffigs{b2ofb1ofp}{bK2ofb1ofp}).

\end{itemize}

Our results on the impact of assembly bias on the $\bkk(b_1)$ relation have interesting and important consequences for the analysis of galaxy survey data. On the one hand, a number of existing cosmological analyses based on galaxy clustering data keep $\bkk(b_1)$ fixed to the LLIMD relation, e.g.~\cite{Sanchez:2016sas,Beutler:2016arn,2020A&A...633L..10T,2021A&A...646A.140H,2021arXiv210513549D}. While this is likely not a serious source of systematics given the constraining power of current galaxy surveys, our results strongly suggest revisiting the adoption of the LLIMD relation in forecast and constraint studies using future galaxy data. For example, should the galaxies of a given observed galaxy sample reside preferentially in more or less concentrated halos, then our results show that any tight, monotonically decreasing $\bkk(b_1)$ relation (like the LLIMD relation) will likely result in a poor approximation to the actual $\bkk$ of the galaxy sample which in turn may bias cosmological results. In other words, should the use of priors on $\bkk(b_1)$ prove necessary in future data analyses, they may need to be significantly wider than previously thought. Indeed, in our companion paper \cite{Barreira:2021ukk}, where we estimated $\bkk$ for simulated galaxies, we did observe similar orthogonal features to those in \reffig{bK2ofb1ofp} (see figures 4-5 there), which supports the hypothesis that the galaxy distribution at least partly inherits the assembly bias signal of the halos. On the other hand, a significant departure in the $\bkk(b_1)$ relation measured for galaxy samples from that seen in mass-selected halos could be a potential hint of assembly bias. We leave a more detailed investigation of this possibility for future work.

Together with our companion paper \cite{Barreira:2021ukk}, our results demonstrate overall the efficiency of forward modeling to estimate and study galaxy/halo bias, and with relatively modest simulation volumes compared to previously reported bias measurements from cross-correlations of halo-matter statistics, or analyses of galaxy power spectra and bispectra, which used appreciably larger total simulation volumes \cite{Lazeyras:2017,Abidi:2018,Oddo:2019run,Eggemeier:2021cam}. In future work, it would be interesting to study more quantitatively the imprints of the assembly bias of dark matter halos on the $\bkk(b_1)$ relation of galaxies. It is also important to include and study the impact of higher-than-third-order terms in the bias expansion, as well as to improve the renormalization procedure to renormalize higher-order operators as well. This will allow for a more precise understanding of the $\Lambda$ dependence of the bias measurements, and hence the small differences of some of our results here compared to previous works. This will also allow us to extend the study (including the impact of assembly bias) to the third-order bias parameters, which enter in analyses of the halo power spectrum and bispectrum starting at 1-loop order. 

\acknowledgments{We thank Beatriz Tucci for comments and suggestions on the paper draft. TL is supported by the INFN INDARK grant. AB acknowledges support from the  Excellence  Cluster  ORIGINS  which  is  funded  by  the  Deutsche  Forschungsgemeinschaft  (DFG, German Research Foundation) under Germany’s Excellence Strategy - EXC-2094-390783311.  FS acknowledges support from the Starting Grant (ERC-2015-STG 678652) “GrInflaGal” of the European Research Council.  The simulations used in this work and their numerical analysis was  done  on  the  Freya  supercomputer  at  the  Max  Planck Computing and Data Facility (MPCDF) in Garching near Munich.}

%%%%%%%%%%%%%%%%%%%%%%%%%%%%%%%%%%%%%%%%%%%%%%%%%%%%%%%%%%%%%%%%%%%%%%%%%%%%%
%%%%%%%%%%%%%%%%%%%%%%%%%%%%%%%%%%%%%%%%%%%%%%%%%%%%%%%%%%%%%%%%%%%%%%%%%%%%%
%\appendix

%%%%%%%%%%%%%%%%%%%%%%%%%%%%%%%%%%%%%%%%%%%%%%%%%%%%%%%%%%%%%%%%%%%%%%%%%%%%
\FloatBarrier
\bibliography{references}
\end{document}